\documentclass[ aps, showpacs, showkeys, nofootinbib, floatfix, superscriptaddress]{revtex4}

\usepackage{amsfonts}

\usepackage{amssymb}
\usepackage{amsmath}
\usepackage{graphicx}

\begin{document}

\title{Cosmology in Poincar$\acute{e}$ gauge gravity with a pseudoscalar torsion}

 \author{Jianbo Lu}
 \email{lvjianbo819@163.com}
 \affiliation{Department of Physics, Liaoning Normal University, Dalian 116029, P. R. China}
 \author{Guoying Chee}
 \affiliation{Department of Physics, Liaoning Normal University, Dalian 116029, P. R. China}
 \email{qgy8475@sina.com}

\begin{abstract}
 A cosmology of Poincar$\acute{e}$ gauge theory is developed, where several properties of universe corresponding to the cosmological equations with the pseudoscalar torsion function are investigated. The cosmological constant is found  to be the intrinsic torsion and curvature of the vacuum universe and is derived from the theory naturally rather than added artificially, i.e. the dark energy originates from geometry and includes the cosmological constant but differs from it. The cosmological constant puzzle, the coincidence and fine tuning problem are relieved naturally at the same time. By solving the cosmological equations, the analytic cosmological solution is obtained and can be compared with the $\Lambda $CDM model. In addition, the expressions of density parameters of the matter and the geometric dark energy are derived, and it is shown that the evolution of state equations for the geometric dark energy agrees with the current observational data. At last, the full equations of linear cosmological perturbations and the solutions are obtained.
\end{abstract}



\maketitle

\section{$\text{Introduction}$}

 { The discovery of the accelerated expansion of the universe motivates a large variety of theoretical works to explain it. In order to account for the acceleration the Einstein equation has to be modified and then two approaches are developed. One is to introduce ''dark energy'' in the right-hand side in the framework of general relativity (see \cite{1} for recent reviews). The another is to modify the left-hand side of the equation, called modified gravitational theories, e.g., $f(R)$ gravity (see \cite{2} for recent reviews). A large amount of literature in every approach has been accumulated in the past years. However, none of them offers a convincing explanation of the observed results, and most of them were introduced to explain the acceleration phenomenologically, rather than emerging naturally out of fundamental physics principles. The most popular model in the fist approach is the $\Lambda$ cold dark matter ($\Lambda $CDM) model which is plagued by the cosmological constant problem and the coincidence and fine tuning problem. Meanwhile, there is not the enough evidence on the validity of this model. It is shown that the thermal and mechanical stability requirement provides an evidence against the dark energy hypothesis \cite{3}. Adding dark energy to the content of the Universe may not be the answer to the cosmic acceleration problem. In the second approach the Einstein-Hilbert Lagrangian is usually generalized to a function $f$ of the Ricci scalar $R$. However, at present there are no fully realized and empirically viable $f$ $\left( R\right) $ theories that explain the observed level of cosmic acceleration. Furthermore, the $f$ $\left( R\right) $ theories suffer from a long-standing controversy about which frame (Einstein or Jordan) is the physical one \cite{4}. It should be noted that although we have strong observational evidence for accelerated cosmic expansion but no compelling evidence that the cause of this acceleration is really a\ new energy component. At the same time we do not have enough independent data yet to clarify the\ nature of dark energy. This provides further motivation for a deeper investigation of the nature of dark
energy or the origin of the accelerated cosmic expansion. In the framework of $f(R)$ gravity, the field equation can be written as the Einstein equation with an effective energy-momentum tensor that contains all the modifications and the energy-momentum tensor of matter fields. The contributions of the modifications of gravity can be identified with some kind of geometric dark energy. This is specially advantageous since one can define an equation of state associated with such dark energy and compare it with the $\Lambda $CDM model \cite{5}. However, the function $f(R)$ is not known a priori, none introduces a new fundamental principle that can be used as a guiding line, it is usually constructed by trial and error. In fact, as a geometric theory a modified gravity should be formulated in a gauge theoretical framework. A famous example is the Poincar$\acute{e}$ gauge theory of Gravity \cite{6}. Some works have been done to develop a model of geometric dark energy in{\bf \ }the{\bf \ }Poincar$\acute{e}$ gauge theory framework \cite{7,8,9}. In \cite{7} the effect of torsion is to introduce an extra-term into matter density and pressure which gives rise to an accelerated behavior of the universe. However, the torsion contributes only a constant density, it is not possible to solve the coincidence and fine tuning problem. The torsion model in \cite{8} contributes an
oscillating aspect to the expansion rate of the universe and displays features similar to those of only the presently observed accelerating universe. In \cite{9} the Lagrangian involves too many terms and indefinite parameters, which make the field equations complicated and difficult to solve and the role of each term obscure. In order to simplify the field equations, some restrictions on indefinite parameters have to be imposed. Under these restrictions, especially, all the higher derivatives of the scale factor are excluded from the cosmological equations.

In fact, starting from a well behaved Lagrangian $\frac 12R+\alpha R^2+\beta R_{\mu \nu }R^{\mu \nu }$ in quadratic gravity \cite{10} and string theory \cite{11} and adding a quadratic term of torsion $\gamma T^\mu {}_{\nu \rho }T_\mu {}^{\nu \rho }$ a good toy model can be obtained \cite{12}, where  we derive to give the gravitational field equation and the cosmological evolutional equations. When the macroscopic spacetime average of the spin vanishes, the cosmological equations are found to split into two families. Each of them is related with only one torsion function, the scalar or the pseudoscalar torsion function. It has been argued \cite{8} that only these two scalar torsion modes are physically acceptable and no-ghosts. This model has a free-ghost dynamics. It has a well posed initial value problem without any ghost or tachyonic propagation. Also, the field equations are allowed to contain higher derivatives in \cite{12}, which is different from  \cite{9}, where some restrictions on indefinite parameters are imposed in order to exclude higher order derivatives. In addition, for the  first family we solve the cosmological equations  corresponding to the scalar torsion function  by using the dynamical system approach in Ref.\cite{12}. In this paper, we study the second-family  cosmological equations corresponding to the pseudoscalar torsion function in detail, with using a totally different way. Some meaningful consequences can be inferred, such as the geometrical interpretation of cosmological constant is investigated, the cosmological constant problem and the coincidence and fine tuning problem are solved naturally, the state equation of the geometrical dark energy is derived and its evolution is consistent with the current observations, the analytic solution $a=a(t)$ in cosmology is obtained and the perturbation analysis is given, etc.

In Sec. II, starting from the Poincar$\acute{e}$ gauge principle and the simple Lagrangian $%
\frac 12R+\alpha R^2+\beta R_{\mu \nu }R^{\mu \nu }+\gamma T^\mu {}_{\nu\rho }T_\mu {}^{\nu \rho }$, which is the sum of the Starobinsky Lagrangian
\cite{13} and Yang-Mills type terms of the local rotation and translation field strength, we introduce the main equations in this  Poincar$\acute{e}$ gauge cosmology, including the gravitational field equations and the cosmological equations, etc. In order to evade any unnecessary discussion regarding frames (i.e. Einstein .vs. Jordan) the theory is treated using the original variables instead of transforming to a scalar-tensor theory in contrast to $f(R)$ theories. Furthermore,  a set of cosmological equations  corresponding to the pseudoscalar torsion function are discussed, where it is found that although we do not introduce a cosmology constant in the action it automatically emerges in the derivation of the cosmological equations and then is endowed with intrinsic character. The dark energy is identified with the geometry of the spacetime and is a function of the density and the pressure of the matter. It includes the cosmological constant\ but can not be identified with it. It is nothing but the intrinsic torsion or curvature of the vacuum universe. In Sec. III, the analytic expressions of the state equation and the density parameters of the matter and the geometric dark energy are derived and used to determine the values of $\alpha $ , $\beta $ and $\gamma $. Then a theoretical value of the cosmological constant is computed and compared with the observed datum. The cosmological constant problem and the coincidence and fine tuning
problem are solved naturally. In Sec. IV an analytic integral of the cosmological equation is obtained and used to evaluate the age of the universe which can be compared with observed data. In Section V the full equations of linear cosmological perturbations and the solutions are obtained. In addition, the behavior of perturbations for the sub-horizon modes relevant to large-scale structures is discussed. It is shown that our model can be distinguished from others by considering the evolution of matter perturbations and gravitational potentials. Sec. VI is devoted to conclusions.

\section{$\text{Cosmological  equations}$}

The discussion in this paper are entirely classical. We consider a Poincar$\acute{e}$
gauge theory of gravity \cite{6,7,8,9}, in which
there are two sets of local gauge potentials, the orthonormal frame field
(tetrad) $e_I{}^\mu $ and the metric-compatible connection $\Gamma
{}^{IJ}{}_\mu $ associated with the translation and the Lorentz subgroups of
the Poincar$\acute{e}$ gauge group, respectively. We use the Greek alphabet ($\mu $, $%
\nu $, $\rho $, $...=0,1,2,3$) to denote (holonomic) indices related to
spacetime, and the Latin alphabet ($I,J,K,...=0,1,2,3$) to denote algebraic
(anholonomic) indices, which are raised and lowered with the Minkowski
metric $\eta _{IJ}$ $=$ diag ($-1,+1,+1,+1$). The field strengths associated
with the tetrad and connection are the torsion $T^\lambda {}_{\mu \nu }$ and
the curvature $R_{\mu \nu }{}^{\lambda \tau }$. We use the geometrized
system of units in which $8\pi G=1$, $c=1$, and start from the action
\begin{equation}
S=\int d^4x\sqrt{-g}\left[ \left( \frac 12R+\alpha R^2+\beta R_{\mu \nu
}R^{\mu \nu }+\gamma T{}^\mu {}_{\nu \rho }T{}_\mu {}^{\nu \rho }\right) +%
{\cal L}_m\right] ,
\end{equation}
where ${\cal L}_m$ denotes the Lagrangian of the source matter including
baryonic matter, cold dark matter and radiation, $\alpha $ and $\beta $ are
two parameters with the dimension of $\left[ L\right] ^2$, $\gamma $ is a
dimensionless parameter. The vales of $\alpha $, $\beta $ and $\gamma $ can
be determined by experiment and observational data.  The terms $\frac 12R$ and
$\gamma T{}^\mu {}_{\nu \rho }T{}_\mu {}^{\nu \rho }$ represent weak
gravity, while $\alpha R^2$ and $\beta R_{\mu \nu }R^{\mu \nu }$ represent
strong gravity with the dimensionless strong gravity constant $\alpha $ and $%
\beta $ according to Hehl {\it et al }\cite{6}.

The variational principle yields the field equations for the tetrad{\em \ } $%
e_I=e_I{}^{\mu}\partial_{\mu} $ and the connection $\Gamma {}^{IJ}=\Gamma {}^{IJ}_{~~\mu}dx^{\mu} $ \cite{6}:
\begin{equation}
DH_{I}-T_{(g)I}=T_{I} (first),\nonumber
\end{equation}
\begin{equation}
DH_{IJ}-s_{(g)IJ}=s_{IJ} (second),\nonumber
\end{equation}
with the covariant derivatives ($D$) of the translation excitation $H_{I}$ and the Lorentz excitation $H_{IJ}$, the gauge currents of
energy-momentum $T_{(g)I}$ and spin $s_{(g)IJ}$ and the canonical matter currents of energy-momentum $T_{I}$ and spin (angular
momentum) $s_{IJ}$. The reduced explicit form of these field equations are [12]:
\begin{equation}
R{}_{\nu \mu }-\frac 12g{}_{\nu \mu }R=T_{\nu \mu }+T_{(g)\nu \mu },
\end{equation}
and
\begin{equation}
T^\nu \!_{\tau \nu }\delta _\lambda ^\mu -T^\nu \!_{\lambda \nu }\delta
_\tau ^\mu +T^\mu \!_{\nu \tau }\delta _\lambda ^\nu =e^I\!_\lambda
e^J\!_\tau \left( s_{IJ}{}^\mu +s_{(g)IJ}{}^\mu \right) ,
\end{equation}
where $T{}_{\nu \mu }:=e_{I\mu }\partial \left( \sqrt{-g}{\cal L}_m\right)
/\partial e_I{}^\nu $ and $s_{IJ}{}^\mu :=\partial \left( \sqrt{-g}{\cal L}%
_m\right) /\partial \Gamma {}^{IJ}{}_\mu $ are energy-momentum tensor and
spin tensor of the source matter, respectively, while
\begin{eqnarray}
T_{(g)\nu \mu } &=&-\alpha \left( 4R{}_{\nu \mu }-g{}_{\nu \mu }R\right)
R-\beta \left( 2R{}^\rho {}_\nu R{}_{\rho \mu }+2R{}^{\rho \sigma
}{}R{}_{\nu \rho \mu \sigma }-g{}_{\nu \mu }R_{\rho \sigma }R^{\rho \sigma
}\right)  \nonumber \\
&&+\gamma \left( 4\partial _\tau e{}^{I\lambda }\left( e_{I\nu }T{}_{\mu
\lambda }{}^\tau -e{}_{I\lambda }T{}_{\mu \nu }{}^\tau \right) +4\partial
_\tau T{}_{\mu \nu }{}^\tau +g{}_{\nu \mu }T{}^\lambda {}_{\rho \sigma
}T{}_\lambda {}^{\rho \sigma }-4T{}^\lambda {}_{\nu \tau }T{}{}_{\lambda \mu
}{}^\tau \right) ,
\end{eqnarray}
and
\begin{eqnarray}
s_{(g)IJ}{}^\mu &=&-4\alpha \left( e_{[I}{}^\nu e_{J]}{}^\tau \Gamma ^\mu
{}_{\nu \tau }R+e_{[J}{}^\mu e_{I]}{}^\nu \left( \Gamma ^\lambda {}_{\lambda
\nu }R-\partial _\nu R\right) +e_{[I}{}^\nu e_{J]}{}^\mu Re{}^K{}_\tau
\partial _\nu e_K{}^\tau \right)  \nonumber \\
&&-4\beta e_J{}^\lambda \left( e_I{}^{[\mu }\partial _\nu R_\lambda {}^{\nu
]}+e_I{}^{[\nu }R_\lambda {}^{\mu ]}e{}^K{}_\tau \partial _\nu e_K{}^\tau
+e_I{}^\tau \Gamma ^{[\nu }{}_{\nu \tau }R_\lambda {}^{\mu ]}+e_I{}^{[\nu
}R_\tau {}^{\mu ]}\Gamma ^\tau {}_{\nu \lambda }\right)  \nonumber \\
&&-4\gamma e_{I\nu }e{}_J{}^\tau T{}^{\nu \mu }{}_\tau ,
\end{eqnarray}
are the energy-momentum and the spin of this kind of ''geometric dark
energy'' corresponding to the terms $\alpha R^2+\beta R_{\mu \nu }R^{\mu \nu
}+\gamma T{}^\mu {}_{\nu \rho }T{}_\mu {}^{\nu \rho }$ in (1). Note that the
energy-momentum tensor $T{}_{\nu \mu }$ of type ($0$, $2$) should not be
confused with the torsion tensor $T^\lambda {}_{\mu \nu }$ of type ($1$, $2$%
). If $\alpha =\beta =\gamma =0$, these equations become the field equations
of Einstein-Cartan-Sciama-Kibble theory. Furthermore, for $T^\lambda {}_{\mu
\nu }=0$ we come back to General Relativity.
 The Eqs. (2-5)  can be rewritten in terms of the covariant derivative.
  Here given that they are convenient in the concrete calculation for deriving the following cosmological equations,
  we exhibit these expressions (2-5) directly.

For the spatially flat Friedmann-Robertson-Walker (FRW) metric
\begin{equation}
g_{\mu \nu }=\text{diag}\left( -1,a\left( t\right) ^2,a\left( t\right)
^2,a\left( t\right) ^2\right) ,
\end{equation}
the non-vanishing torsion components with holonomic indices are given by two
functions, the scalar torsion $h$ and the pseudoscalar torsion $f$ \cite{8,14}:
\begin{equation}
T_{ij0}=a^2h\delta _{ij},T_{ijk}=2a^3f\epsilon
_{ijk},\;\;\;\;i,j,k,...=1,2,3.
\end{equation}
The equations (2) and (3) yields the cosmological equations \cite{12}\footnote{It can be seen that the symbol of $p$ (denotes pressure) in Eq.(9) is opposite to the one in Eq.(15) of Ref.[12]. The reason is that there is a error on symbol of $p$ in [12]. The wrong symbol on $p$ in Eq.(15) of Ref.[12] results that $¡°p¡±$ (or $\dot{p}$ or $w=p/\rho$) in all formulas of Ref.[12] should be corrected to $¡°-p¡±$ (or $-\dot{p}$ or $-w $). However, the main results and conclusions of Ref.[12] remain unaffected, since the main discussions are performed in vacuum universe in Ref.[12]. This wrong sign of $p$ has been corrected in this paper. Also, the corrected equations can be found in the errata regarding Ref.[12]. }

\begin{equation}
H^2=\frac 13\left( \rho +\rho _g\right) ,
\end{equation}
\begin{equation}
2\stackrel{\cdot }{H}+3H^2=-\left( p+p_g\right) ,
\end{equation}
\begin{eqnarray}
&&\left( \beta +6\alpha \right) \left( \stackrel{\cdot \cdot }{H}+\stackrel{%
\cdot \cdot }{h}\right) +6\left( \beta +4\alpha \right) \left( H+h\right)
\stackrel{\cdot }{H}+\left( \allowbreak 5\beta +18\alpha \right) \left(
H+h\right) \stackrel{\cdot }{h}-4\left( \beta +3\alpha \right) f\stackrel%
{\bf \cdot }{f}  \nonumber \\
&&+3\left( \beta +4\alpha \right) hH^2+\left( \allowbreak 5\beta +18\alpha
\right) h^2H+2\left( \beta +3\alpha \right) h^3-2\left( \beta +3\alpha
\right) hf^2+\frac 14h+\frac 12s_{01}{}^1=0,
\end{eqnarray}
and
\begin{eqnarray}
&&f\{2\left( \beta +6\alpha \right) \left( \stackrel{\cdot }{H}+\stackrel{%
\cdot }{h}\right) +6\left( \beta +\allowbreak 4\alpha \right) H^2+2\left(
5\beta +18\alpha \right) Hh  \nonumber \\
&&+\left( \beta \ +3\alpha \right) \left( 4h^2-4f^2\right) -4\gamma +\frac 12%
\}-\frac 12s_{12}{}^3=0,
\end{eqnarray}
where $H=\stackrel{\cdot }{a}\left( t\right) /a\left( t\right) $ is the
Hubble parameter, $\stackrel{\cdot }{H}=dH/dt$, while
\begin{eqnarray}
\rho _g &=&-6Hh-3h^2+3f^2  \nonumber \\
&&+12\left( 3\alpha +\beta \right) \left( \stackrel{\cdot }{H}+\stackrel{%
\cdot }{h}-Hh-h^2+f^2\right) \left( \stackrel{\cdot }{H}+\stackrel{\cdot }{h}%
+2H^2+3Hh+h^2-f^2\right)   \nonumber \\
&&-2\gamma \left( {}3h^2+4f^2\right)
,\;\;\;\;\;\;\;\;\;\;\;\;\;\;\;\;\;\;\;\;\;\;\;\;\;\;\;\;\;\;\;\;\;\;
\end{eqnarray}
and
\begin{eqnarray}
p_g &=&4{}Hh+{}h^2-f^2+2\dot{h}  \nonumber \\
&&+4\left( 3\alpha +\beta \right) \left( \stackrel{\cdot }{H}+\stackrel{%
\cdot }{h}-Hh-h^2+f^2\right) \left( \stackrel{\cdot }{H}+\stackrel{\cdot }{h}%
+2H^2+3Hh+h^2-f^2\right)   \nonumber \\
&&-2\gamma \left( 2\stackrel{\cdot }{h}+8Hh+h^2+4f^2\right)
,\;\;\;\;\;\;\;\;\;\;i=1,2,3,\;\;\;\;\;\;\;\;\;\;\;\;\;\;\;
\end{eqnarray}
are the density and the pressure of the geometric dark energy. (12) and (13)
indicate that the geometric dark energy is just the gravitational field
itself described by $h$, $H$ and $f$. The source matter is a fluid
characterized by the energy density $\rho =T_{00}$, the pressure $p=T_{ij}$ (%
$i=j$) and the spin $s_{IJ}{}^\mu $. (8) and (9) lead to
\begin{equation}
\frac{\stackrel{\cdot \cdot }{a}}a=-\frac 16\left( \rho +\rho
_g+3p+3p_g\right) .
\end{equation}
It is easy to see that when $\alpha =\beta =\gamma =0$ and $h=f=0$, (8), (9) and
(14) reduce to the Friedmann cosmology. (8) and (9) correspond to the
Friedmann equation and the Raychaudhuri equation respectively, while (14) is
the acceleration equation, which represent the Einstein frame of the theory.

Since the spin orientation of particles for ordinary matter is random, the
macroscopic spacetime average of the spin vanishes, we suppose $%
s_{IJ}{}^\lambda =0$, henceforth. Then, the equation (11) has the solutions
\begin{equation}
f=0,
\end{equation}
and
\begin{eqnarray}
f^2 &=&\frac{\left( \beta +6\alpha \right) }{2\left( \beta \ +3\alpha
\right) }\left( \stackrel{\cdot }{H}+\stackrel{\cdot }{h}\right) +\frac{%
3\left( \beta +\allowbreak 4\alpha \right) }{2\left( \beta \ +3\alpha
\right) }H^2+\frac{\left( 5\beta +18\alpha \right) }{2\left( \beta \
+3\alpha \right) }Hh+h^2  \nonumber \\
&&-\frac \gamma {\left( \beta \ +3\alpha \right) }+\frac 1{8\left( \beta \
+3\alpha \right) }.
\end{eqnarray}

The solution (15) has been investigated in \cite{12}. We concentrate on the
equation (16) now. Differentiating (16) gives

\[
f\stackrel{\cdot }{f}=\frac{\beta +6\alpha }{4\left( \beta +3\alpha \right) }%
\left( \stackrel{\cdot \cdot }{H}+\stackrel{\cdot \cdot }{h}\right) +\frac{%
3\left( \beta +\allowbreak 4\alpha \right) }{2\left( \beta +3\alpha \right) }%
H\stackrel{\cdot }{H}+\frac{5\beta +18\alpha }{4\left( \beta +3\alpha
\right) }\stackrel{\cdot }{H}h+\frac{5\beta +18\alpha }{4\left( \beta
+3\alpha \right) }H\stackrel{\cdot }{h}+h\stackrel{\cdot }{h}.
\]
Substituting it and (16) into (10) gives (when $s_{01}{}^1=0$)

\[
2h\gamma =0,
\]
and then

\begin{equation}
h=0.
\end{equation}
In this case (8), (9), (12), (13) and (16) lead to

\[
12\left( 3\alpha +\beta \right) \left( \stackrel{\cdot }{H}+f^2\right)
\left( \stackrel{\cdot }{H}+2H^2-f^2\right) -3H^2+3f^2-24\gamma f^2+\rho =0,
\]
\[
4\left( 3\alpha +\beta \right) \left( \stackrel{\cdot }{H}+f^2\right) \left(
\stackrel{\cdot }{H}+2H^2-f^2\right) +2\stackrel{\cdot }{H}+3H^2-f^2-8\gamma
f^2+p=0,
\]
\[
\left( \beta +6\alpha \right) \stackrel{\cdot }{H}+3\left( \beta
+\allowbreak 4\alpha \right) H^2-2\left( \beta \ +3\alpha \right)
f^2-2\gamma +\frac 14=0,
\]
which further give$\allowbreak $
\begin{equation}
f^2=\frac \Lambda {24\gamma }+\frac \rho {24\gamma }+\allowbreak \frac{%
3\left( 4\alpha +\beta \right) -16\gamma \left( 3\alpha +\beta \right) }{%
48\beta \gamma }\left( \rho -3p\right) +\allowbreak \frac{\left( 3\alpha
+\beta \right) \left( 4\alpha +\beta \right) }{24\gamma \beta }\left( \rho
-3p\right) ^2,
\end{equation}
\begin{equation}
H^2=\frac{\left( 1-8\gamma \right) }{24\gamma }\Lambda +\frac \rho {24\gamma
}-\allowbreak \frac{\left( 8\gamma -1\right) \left( 4\alpha +\beta \right) }{%
16\gamma \beta }\left( \rho -3p\right) +\allowbreak \frac{\left( 3\alpha
+\beta \right) \left( 4\alpha +\beta \right) }{24\gamma \beta }\left( \rho
-3p\right) ^2,
\end{equation}
$\allowbreak $
\begin{eqnarray}
\stackrel{\cdot }{H} &=&\allowbreak \frac{16\gamma -1}{24\gamma }\Lambda -%
\frac \rho {24\gamma }\allowbreak   \nonumber \\
&&-\frac{3\left( 4\alpha +\beta \right) -8\gamma \left( 18\alpha +5\beta
\right) }{48\beta \gamma }\left( \rho -3p\right) \allowbreak -\allowbreak
\frac{\left( 3\alpha +\beta \right) \left( 4\alpha +\beta \right) }{24\gamma
\beta }\left( \rho -3p\right) ^2\allowbreak ,
\end{eqnarray}
where
\begin{equation}
\Lambda =\frac{3\left( 1-8\gamma \right) }{4\beta },
\end{equation}
is the geometric cosmological constant coming from the terms $\beta R_{\mu \nu
}R^{\mu \nu }+\gamma T{}^\mu {}_{\nu \rho }T{}_\mu {}^{\nu \rho }$ in (1).
It is easy to see that the higher order derivative $\stackrel{\cdot \cdot }{H}$ and $%
\stackrel{\cdot }{f}$ in (10) disappear. We also note that although we do not introduce
a cosmological constant $\Lambda $ in the action (1), it automatically emerges
in these equations. In (18) the pseudoscalar torsion $f$ is a function of $%
\rho $ and $p$ rather than a constant, in contrast to \cite{5}. It should be
noted that although (8) and (9) have the same form as the Friedmann
equations, the solutions (19) and (20) are different. The reason is that in
(8) and (9) $\rho _g$ and $p_g$ are functions of $H$, $\stackrel{\cdot }{H}$%
, $h$, $\stackrel{\cdot }{h}$, and $f$ as indicated by (12) an (13). In
other words, this is a different model from the $\Lambda $CDM model
essentially.

(12) and (13) become now
\begin{eqnarray}
\rho _g &=&\frac{\left( 1-8\gamma \right) }{8\gamma }\Lambda +\frac{\left(
1-8\gamma \right) \rho }{8\gamma }\allowbreak +\frac{3\left( 1-8\gamma
\right) \left( 4\alpha +\beta \right) }{16\beta \gamma }\left( \rho
-3p\right) +\allowbreak \frac{\left( 3\alpha +\beta \right) \left( 4\alpha
+\beta \right) }{8\beta \gamma }\left( \rho -3p\right) ^2, \\
p_g &=&-\frac{\left( 1+8\gamma \right) }{24\gamma }\Lambda -\frac{\left(
8\gamma +1\right) \rho }{24\gamma }-\allowbreak \frac{3\left( 4\alpha +\beta
\right) -8\beta \gamma }{48\beta \gamma }\left( \rho -3p\right) -\frac{%
\left( 3\alpha +\beta \right) \left( 4\alpha +\beta \right) }{24\beta \gamma
}\left( \rho -3p\right) ^2,
\end{eqnarray}
which mean that  {\it the geometrical dark energy includes the
cosmological constant }$\Lambda ${\it \ but can not be identified with it. }%
The cosmological constant $\Lambda $ is really a constant determined by $%
\beta $ and $\gamma $ as indicated by (20) while the geometrical dark energy $\rho _g$
is a function of the density $\rho $ and the pressure $p$ of the matter.
{\it The cosmological constant problem and the coincidence and fine tuning
problem are relieved naturally, as shown in the next section.}

Substituting (22) and (23) into (14) yields

\begin{equation}
\frac{\stackrel{\cdot \cdot }{a}}a=\frac \Lambda 3+\frac{3\alpha +\beta }{%
3\beta }\left( \rho -3p\right) .
\end{equation}
Furthermore, (18), (19), (20) and (24) mean that the vacuum universe has the
torsion
\begin{equation}
f_{\text{vac}}^2=\frac \Lambda {24\gamma },
\end{equation}
the curvature
\begin{equation}
R_{\text{vac}}=6\stackrel{\cdot }{H}+12H^2-3f^2=\allowbreak \frac \Lambda {%
8\gamma }.
\end{equation}
and the acceleration
\begin{equation}
\left( \frac{\stackrel{\cdot \cdot }{a}}a\right) _{\text{vac}}=\frac \Lambda %
3.
\end{equation}
This means that the cosmological constant is nothing but the intrinsic
torsion or curvature of the vacuum universe.


\section{$\text{The state equation of the geometrical dark energy}$}

(22) and (23) gives the state equation of the dark energy:
\begin{equation}
w_g=\frac{p_g}{\rho _g}=\frac{-2\left( 1+8\gamma \right) \beta \Lambda
-2\left( 8\gamma +1\right) \beta \rho -\allowbreak \left( 3\left( 4\alpha
+\beta \right) -8\beta \gamma \right) \left( \rho -3p\right) -2\left(
3\alpha +\beta \right) \left( 4\alpha +\beta \right) \left( \rho -3p\right)
^2}{6\left( 1-8\gamma \right) \beta \Lambda +6\left( 1-8\gamma \right)
\allowbreak \beta \rho +9\left( 1-8\gamma \right) \left( 4\alpha +\beta
\right) \left( \rho -3p\right) +6\left( 3\alpha +\beta \right) \left(
4\alpha +\beta \right) \left( \rho -3p\right) ^2}.
\end{equation}
The source matter includes ordinary baryon matter,  dark matter and radiation:
\begin{equation}
\rho =\rho _{\text{m}}+\rho _{\text{r}},p_{\text{m}}=0,p=\frac 13\rho _{_{%
\text{r}}}.
\end{equation}
The equation (8) can be written as
\[
\Omega =\Omega _{\text{r}}+\Omega _{\text{m}}+\Omega _g=1,
\]
where
\begin{equation}
\Omega _{\text{m}}:=\frac{\rho _{\text{m}}}{3H^2},\Omega _{\text{r}}:=\frac{%
\rho _{\text{r}}}{3H^2},\Omega _g:=\frac{\rho _g}{3H^2},
\end{equation}
are the dimensionless density parameters of the matter, the radiation and
the geometrical dark energy, respectively.

Suppose
\begin{equation}
\alpha =-\frac \beta 2.
\end{equation}
(22), (23) and (24) become
\begin{eqnarray}
\rho _g &=&\frac{1-8\gamma }{8\gamma }\Lambda +\frac{\left( 1-8\gamma
\right) \rho _{_{\text{r}}}}{8\gamma }\allowbreak -\frac{1-8\gamma }{%
16\gamma }\rho _{\text{m}}+\frac \beta {16\gamma }\rho _{\text{m}}^2, \\
p_g &=&-\frac{1+8\gamma }{24\gamma }\Lambda -\frac{\left( 1+8\gamma \right)
\rho _{_{\text{r}}}}{24\gamma }+\frac{1-8\gamma }{48\gamma }\rho _{\text{m}}-%
\frac \beta {48\gamma }\rho _{\text{m}}^2,
\end{eqnarray}
\begin{equation}
w_g=\frac{p_g}{\rho _g}=\frac{-2\left( 1+8\gamma \right) \Lambda -2\left(
1+8\gamma \right) \rho _{_{\text{r}}}+\left( 1-8\gamma \right) \rho _{\text{m%
}}-\beta \rho _{\text{m}}^2}{6\left( 1-8\gamma \right) \Lambda +6\left(
1-8\gamma \right) \rho _{_{\text{r}}}\allowbreak -3\left( 1-8\gamma \right)
\rho _{\text{m}}+3\beta \rho _{\text{m}}^2},
\end{equation}
and
\begin{equation}
\frac{\stackrel{\cdot \cdot }{a}}a=\frac \Lambda 3-\frac 16\left( \rho
-3p\right) =\frac \Lambda 3-\frac 16\rho _{\text{m}}.
\end{equation}
(30) and (19) give
\begin{eqnarray}
\Omega _{\text{m}} &=&\frac{16\gamma \rho _{\text{m}}}{2\left( 1-8\gamma
\right) \Lambda +2\rho _{\text{r}}+\left( 24\gamma -1\right) \rho _{\text{m}%
}+\beta \rho _{\text{m}}^2}, \\
\Omega _g &=&\frac{2\left( 1-8\gamma \right) \Lambda +2\left( 1-8\gamma
\right) \rho _{\text{r}}\allowbreak -\left( 1-8\gamma \right) \rho _{\text{m}%
}+\beta \rho _{\text{m}}^2}{2\left( 1-8\gamma \right) \Lambda +2\rho _{\text{%
r}}+\left( 24\gamma -1\right) \rho _{\text{m}}+\beta \rho _{\text{m}}^2}.
\end{eqnarray}

From the observed data
\[
\rho _{\text{crit}}=1.88h^2\times 10^{-29}gcm^{-3}=7.\,2402\times
10^{-58}cm^{-2},
\]
\begin{eqnarray}
\Omega _{\text{m}} &=&0.3,\Omega _{\text{r}}=1.\,8035\times 10^{-4}\Omega _{%
\text{m}} \\
w_g &=&-1,
\end{eqnarray}
using (31), (34) and (36) we can determine the parameters
\begin{eqnarray}
\alpha  &=&-4.\,1969\times 10^{56}cm^2,  \nonumber \\
\beta  &=&8.\,3937\times 10^{56}cm^2,  \nonumber \\
\gamma  &=&0.0\,576.
\end{eqnarray}
Then (21), (25), (26) and (27) give
\begin{equation}
\Lambda =4.\,8179\times 10^{-58}cm^{-2},
\end{equation}
\begin{equation}
f_{\text{vac}}^2=3.\,4852\times 10^{-58}cm^{-2},
\end{equation}
\begin{equation}
R_{\text{vac}}=1.\,0456\times 10^{-57}cm^{-2},
\end{equation}
and
\begin{equation}
\left( \frac{\stackrel{\cdot \cdot }{a}}a\right) _{\text{vac}}=1.\,606\times
10^{-58}cm^{-2}=1.\,4425\times 10^{-37}s^{-2}.
\end{equation}
The value given by (41) can be compared with the observed datum
\[
\Lambda ^{\left( \text{obs}\right) }=8\pi G\rho _\Lambda ^{\left( \text{obs}%
\right) }=8\pi G\left( 10^{-12}GeV\right) ^4\sim 8\pi G\times 2\times
10^{-10}erg/cm^3=4.\,1574\times 10^{-58}cm^{-2}.
\]

Since

\[
\rho _{\text{r}}=\frac{\overline{\rho }_{\text{r}}}{a^4},\overline{\rho }_{%
\text{r}}=\rho _{\text{r,}a=1}\text{,}\rho _{\text{m}}=\frac{\overline{\rho }%
_{\text{m}}}{a^3},\overline{\rho }_{\text{m}}=\rho _{\text{m,}a=1},
\]
using (40) and
\[
\overline{\rho }_{\text{m}}=0.3\rho _{\text{crit}},\overline{\rho }_{\text{r}%
}=1.\,8035\times 10^{-4}\overline{\rho }_{\text{m}}
\]
the state equation of the dark energy (34) can be written as
\begin{equation}
w_g\left( a\right) =\frac{p_g}{\rho _g}=\frac{-0.\,78766-6.\,4044\times
10^{-5}a^{-4}+6.\,5538\times 10^{-2}a^{-3}-2.216\times 10^{-2}a^{-6}}{%
0.\,87221+7.\,0919\times 10^{-5}a^{-4}\allowbreak
-0.\,19661a^{-3}+6.\,6481\times 10^{-2}a^{-6}},
\end{equation}
or
\begin{equation}
w_g\left( z\right) =\frac{-0.\,78766-6.\,4044\times 10^{-5}\left( 1+z\right)
^4+6.\,5538\times 10^{-2}\left( 1+z\right) ^3-\,2.216\times 10^{-2}\left(
1+z\right) ^6}{0.\,87221\allowbreak +7.\,0919\times 10^{-5}\left( 1+z\right)
^4-0.\,19661\left( 1+z\right) ^3+6.\,6481\times 10^{-2}\left( 1+z\right) ^6}.
\end{equation}
Figure 1 plots the evolution history of $w_{g}(a)$ given by (45).

\begin{figure}[ht]
 \includegraphics[width=5cm]{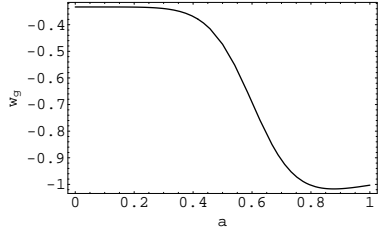}\\
 \caption{ The evolution of $w_{g}(a)$.}\label{wg-a}
\end{figure}

In observation and experiments it is conventional to phrase constraints or
projected constraints on $w(z)$ in terms of a linear evolutional model \cite
{15}:

\[
w(a)=w_0+w_a(1-a)=w_{\text{p}}+w_a\left( a_{\text{p}}-a\right) ,
\]
where $w_0$ is the value of $w$ at $z=0$ ($a=1$), and $w_{\text{p}}$ is the
value of $w$ at a ''pivot'' redshift $z_{\text{p}}$. For typical data
combinations, $z_{\text{p}}\approx 0.5$. To this end we give the linear
approximation of (45). When $a=1$,
\[
w_{g0}=-1,
\]
and
\begin{equation}
\frac{dw_g}{da}|_{a=1}=0.\,19936,
\end{equation}
so we have
\begin{equation}
w_g\left( a\right) =w_{g0}+\frac{dw_g}{da}|_{a=1}\left( a-1\right)
=-1-0.\,19936\left( 1-a\right) .
\end{equation}
When $z_{\text{p}}=0.5$, $a=\frac 23$,
\begin{equation}
w_{g\text{p}}=-0.\,84781,
\end{equation}
and
\begin{equation}
\frac{dw_g}{da}|_{a=\frac 23}=-0.\,7653,
\end{equation}
then we have
\begin{equation}
w_{g\text{p}}\left( a\right) =w_{g\text{p}}+\frac{dw_g}{da}|_{\text{p}%
}\left( a-\frac 23\right) =-0.\,8478+0.\,7653\left( \frac 23-a\right) .
\end{equation}
Using (35) and $\rho _{\text{m}}=\overline{\rho }_{\text{m}}/a^3$, one finds
that when
\begin{equation}
a=a_{\text{trans}}=\left( \frac{2\beta \overline{\rho }_{\text{m}}}{3\left(
1-8\gamma \right) }\right) ^{\frac 13}=0.\,60859,
\end{equation}
\begin{equation}
z_{\text{trans}}=0.64314,
\end{equation}
the expansion of the universe transforms from deceleration to acceleration.
Using (37) one can compute that when

\begin{equation}
a=0.\,75817,z=0.\,31897,\Omega _g=0.5,
\end{equation}
the universe transforms from the matter dominating phase into the dark
energy dominating phase. In a flat $\Lambda $CDM universe with ( $\Omega _m$%
, $\Omega _\Lambda $) = ($0.3$, $0.7$) acceleration begins at $z=0.67$,
while dark energy doesn't dominate the energy density of the universe until $%
z=0.33$ \cite{16}.

\section{$\text{An exact analytic solution of cosmological equation}$}

The equation (19) can be solved in two cases as follows. In the radiation
dominated era
\[
\rho _{\text{r}}=\frac{\overline{\rho }_{\text{r}}}{a^4},\overline{\rho }_{%
\text{r}}=\rho _{\text{r,}a=1}=\text{const, }p_{\text{r}}=\frac 13\rho _{%
\text{r}},
\]
(19) reads
\begin{equation}
H^2=\frac{\left( 1-8\gamma \right) ^2}{32\gamma \beta }+\frac{\overline{\rho
}_{\text{r}}}{24\gamma a^4},
\end{equation}
and can be rewritten
\begin{equation}
\frac{da}{dt}=a\sqrt{\frac{\left( 1-8\gamma \right) ^2}{32\gamma \beta }+%
\frac{\overline{\rho }_{\text{r}}}{24\gamma a^4}}.
\end{equation}
Its integration gives
\begin{equation}
\ln \left( a^2+\sqrt{a^4+\frac{4\beta \overline{\rho }_{\text{r}}}{3\left(
1-8\gamma \right) ^2}}\right) -\ln \sqrt{\frac{4\beta \overline{\rho }_{%
\text{r}}}{3\left( 1-8\gamma \right) ^2}}=\frac{\left( 1-8\gamma \right) }{2%
\sqrt{2\gamma \beta }}t,
\end{equation}
or
\begin{eqnarray}
a &=&\left( \frac{\beta \overline{\rho }_{\text{r}}}{3\left( 1-8\gamma
\right) ^2}\right) ^{\frac 14}\sqrt{\left( e^{\frac{\left( 1-8\gamma \right)
}{2\sqrt{2\gamma \beta }}t}-e^{-\frac{\left( 1-8\gamma \right) }{2\sqrt{%
2\gamma \beta }}t}\right) }  \nonumber \\
&=&\left( \frac{\beta \overline{\rho }_{\text{r}}}{3\left( 1-8\gamma \right)
^2}\right) ^{\frac 14}\sqrt{2\sinh \frac{\left( 1-8\gamma \right) }{2\sqrt{%
2\gamma \beta }}t}.
\end{eqnarray}

In the matter dominated era
\[
\rho _{\text{m}}=\frac{\overline{\rho }_{\text{m}}}{a^3},\overline{\rho }_{%
\text{m}}=\rho _{\text{m,}a=1}=\text{const, }p=0,
\]
(19) reads
\begin{equation}
H^2=\frac{\left( 1-8\gamma \right) ^2}{32\gamma \beta }+\frac{12\alpha
+5\beta -24\gamma \left( 4\alpha +\beta \right) }{48\gamma \beta }\frac{%
\overline{\rho }_{\text{m}}}{a^3}+\allowbreak \frac{\left( 3\alpha +\beta
\right) \left( 4\alpha +\beta \right) }{24\gamma \beta }\frac{\overline{\rho
}_{\text{m}}^2}{a^6},
\end{equation}
and then
\begin{equation}
\frac{da}{dt}=a\sqrt{\frac{\left( 1-8\gamma \right) ^2}{32\gamma \beta }+%
\frac{12\alpha +5\beta -24\gamma \left( 4\alpha +\beta \right) }{48\gamma
\beta }\frac{\overline{\rho }_{\text{m}}}{a^3}+\allowbreak \frac{\left(
3\alpha +\beta \right) \left( 4\alpha +\beta \right) }{24\gamma \beta }\frac{%
\overline{\rho }_{\text{m}}^2}{a^6}}.
\end{equation}
Its integration gives
\begin{eqnarray}
&&\ln \left( \sqrt{a^6+2\frac{12\alpha +5\beta -24\gamma \left( 4\alpha
+\beta \right) }{3\left( 1-8\gamma \right) ^2}\overline{\rho }_{\text{m}%
}a^3+\allowbreak \frac{4\left( 3\alpha +\beta \right) \left( 4\alpha +\beta
\right) }{3\left( 1-8\gamma \right) ^2}\overline{\rho }_{\text{m}}^2}+a^3+%
\frac{12\alpha +5\beta -24\gamma \left( 4\alpha +\beta \right) }{3\left(
1-8\gamma \right) ^2}\overline{\rho }_{\text{m}}\right)   \nonumber \\
&&-\ln \left( \sqrt{\allowbreak \frac{4\left( 3\alpha +\beta \right) \left(
4\alpha +\beta \right) }{3\left( 1-8\gamma \right) ^2}\overline{\rho }_{%
\text{m}}^2}+\frac{12\alpha +5\beta -24\gamma \left( 4\alpha +\beta \right)
}{3\left( 1-8\gamma \right) ^2}\overline{\rho }_{\text{m}}\right)   \nonumber
\\
&=&3\frac{\left( 1-8\gamma \right) }{\sqrt{32\gamma \beta }}t,
\end{eqnarray}
and then
\begin{eqnarray}
&&\allowbreak a=\{\frac 12\left( \sqrt{\allowbreak \frac{4\left( 3\alpha
+\beta \right) \left( 4\alpha +\beta \right) }{3\left( 1-8\gamma \right) ^2}%
\overline{\rho }_{\text{m}}^2}+\frac{12\alpha +5\beta -24\gamma \left(
4\alpha +\beta \right) }{3\left( 1-8\gamma \right) ^2}\overline{\rho }_{%
\text{m}}\right) e^{\frac{3\left( 1-8\gamma \right) }{4\sqrt{2\gamma \beta }}%
t}  \nonumber \\
&&-\frac{12\alpha +5\beta -24\gamma \left( 4\alpha +\beta \right) }{3\left(
1-8\gamma \right) ^2}\overline{\rho }_{\text{m}}  \nonumber \\
&&+\frac{36\alpha +13\beta -48\gamma \left( 1+4\gamma \right) \left( 4\alpha
+\beta \right) }{18\left( 1-8\gamma \right) ^4}\beta \overline{\rho }_{\text{%
m}}\left( \sqrt{\allowbreak \frac{4\left( 3\alpha +\beta \right) \left(
4\alpha +\beta \right) }{3\left( 1-8\gamma \right) ^2}}+\frac{12\alpha
+5\beta -24\gamma \left( 4\alpha +\beta \right) }{3\left( 1-8\gamma \right)
^2}\right) ^{-1}e^{-\frac{3\left( 1-8\gamma \right) }{4\sqrt{2\gamma \beta }}%
t}\}^{\frac 13}.
\end{eqnarray}

In the case (40), equations (58) and (62) become
\begin{equation}
a=0.\,67617\sqrt{2\sinh 2.\,7417\times 10^{-29}t},
\end{equation}
and
\begin{equation}
\allowbreak a=\left( 0.\,17801e^{4.\,1125\times 10^{-29}t}-9.\,8074\times
10^{-2}e^{-4.\,1125\times 10^{-29}t}-7.\,9934\times 10^{-2}\right) ^{\frac 13%
},
\end{equation}
where the time $t$ is in cm. This is a new exact analytic cosmological
solution which resembles but differs from the $\Lambda $CDM solution \cite{17}.


Now we can evaluate the age of the universe using (40), (57) and (61). In
the radiation dominating era $z\gtrsim 3000$ \cite{17}, we have the equation
\[
\ln \left( a^2+\sqrt{a^4+\frac{4\beta \overline{\rho }}{3\left( 1-8\gamma
\right) ^2}}\right) -\ln \sqrt{\frac{4\beta \overline{\rho }}{3\left(
1-8\gamma \right) ^2}}=\frac{\left( 1-8\gamma \right) }{2\sqrt{2\gamma \beta
}}t.
\]
Choosing $z=3000$, then $a=1/3001$, this equation gives
\begin{equation}
t=3.\,298\times 10^{23}cm=3.\,4895\times 10^5Years.
\end{equation}
In the matter dominating era $z\lesssim 3000$, we have the equation

\begin{eqnarray*}
&&\ln \left( \sqrt{1+2\frac{\left( 24\gamma -1\right) }{3\left( 1-8\gamma
\right) ^2}\beta \overline{\rho }a_2^3+\allowbreak \frac 2{3\left( 1-8\gamma
\right) ^2}\left( \beta \overline{\rho }\right) ^2}+a_2^3+\frac{\left(
24\gamma -1\right) }{3\left( 1-8\gamma \right) ^2}\beta \overline{\rho }%
\right)  \\
&&-\ln \left( \sqrt{1+2\frac{\left( 24\gamma -1\right) }{3\left( 1-8\gamma
\right) ^2}\beta \overline{\rho }a_1^3+\allowbreak \frac 2{3\left( 1-8\gamma
\right) ^2}\left( \beta \overline{\rho }\right) ^2}+a_1^3+\frac{\left(
24\gamma -1\right) }{3\left( 1-8\gamma \right) ^2}\beta \overline{\rho }%
\right)  \\
&=&=3\frac{\left( 1-8\gamma \right) }{\sqrt{32\gamma \beta }}\left(
t_2-t_1\right) ,
\end{eqnarray*}
Choosing
\begin{eqnarray*}
z_1 &=&3000,a_1=\frac 1{3001}, \\
z_2 &=&0,a_2=1,
\end{eqnarray*}
we have
\begin{equation}
t_2-t_1=1.\,6383\times 10^{28}cm=1.\,7334\times 10^{10}Years=17.334Gy.
\end{equation}

\section{$\text{Perturbation theory}$}

In order to discriminate lots of dark energy models, it is interested to
seek the additional information other than the background expansion history
of the universe \cite{18}. Now we discuss the dynamics of linear
perturbations and the structure growth of universe.

\subsection{Cosmological perturbations of gravitational potentials and the
torsion}

The perturbed equations can be derived by straightforward and tedious
calculations, following the approach of \cite{19}. The computer software
Maple has been applied to work out the lengthy calculations. We focus on the
scalar perturbations, since they are\ sufficient to reveal the\ basic
features of the theory, allowing for a discussion of the\ growth of matter
overdensities. The perturbed vierbein reads
\begin{eqnarray}
e_\mu ^0 &=&\delta _\mu ^0\left( 1+\phi \right) ,e_\mu ^a=a\delta _\mu
^a\left( 1-\psi \right) ,  \nonumber \\
e_0^\mu  &=&\delta _0^\mu \left( 1-\phi \right) ,e_a^\mu =\frac 1a\delta
_a^\mu \left( 1+\psi \right) .
\end{eqnarray}
in which we have introduced the scalar modes $\phi $ and $\psi $ as
functions of $t$. This induces a metric perturbation of the known form,
namely

\begin{equation}
ds^2=a^2(\eta )[-(1+2\phi )d\eta ^2+(1-2\psi )\gamma _{ij}dx^idx^j],\;\;
\end{equation}
in the longitudinal gauge and the conformal time $\eta $.

In order to preserve the global homogeneity and isotropy of the spacetime
the perturbations are assumed to be small. It has been argued \cite{8} that
only two scalar torsion modes $h$ and $f$ are physically acceptable and
no-ghosts. On the basis of the above theoretical tests (e.g., ''no-ghosts'' or
''no-tachyons''), we use (7) to give the linear perturbation of the
nonvanishing torsion components

\[
\delta T_{ij0}=\delta _{ij}a^2\delta h,\delta T_{ijk}=2\epsilon
_{ijk}a^3\delta f,\;\;\;\;i,j,k,...=1,2,3.
\]
In the case (16), $h=0$, we have
\begin{equation}
\delta T_{ij0}=0,\delta T_{ijk}=2\epsilon _{ijk}a^3\xi
,\;\;\;\;i,j,k,...=1,2,3.
\end{equation}
where$\;\xi =\delta f$.

The unperturbed field equation (2) can be written as
\[
G^\mu {}_\nu =T^\mu {}_\nu +T_{(g)}{}^\mu {}_\nu ,
\]
where $G^\mu {}_\nu $ is the Einstein tensor, $T^\mu {}_\nu $ is the
energy-momentum of the ordinary matter and the radiation, $T_{(g)}{}^\mu
{}_\nu $ is the energy-momentum of the ''geometric dark energy'' given by
(4). The equations of motion for small perturbations linearized on the
background metric are
\begin{equation}
\delta G^\mu {}_\nu =\delta T^\mu {}_\nu +\delta T_{(g)}{}^\mu {}_\nu .
\end{equation}
For scalar type metric perturbations with a line element given in (68) (in
conformal time), the perturbed field equations can be obtained following the
approach of \cite{19}.

The cosmic fluid includes radiation, baryonic matter and dark matter, $\rho _{%
\text{m}}=\rho _b+\rho _d$, we have
\begin{equation}
\rho =\rho _{\text{m}}+\rho _{\text{r}}=\rho _b+\rho _d+\rho _r,p=p_r=\frac 1%
3\rho _r.
\end{equation}
Since
\begin{equation}
\rho _b\propto \frac 1{a^3},\rho _r\propto \frac 1{a^4},
\end{equation}
we suppose
\begin{equation}
\rho _d\propto \frac 1{a^n}.
\end{equation}
Then we have
\begin{equation}
\rho _r=r\rho _b,\rho _d=v\rho _b,
\end{equation}
where
\begin{equation}
r\propto a^{-1},v\propto a^{3-n}.
\end{equation}
The equation
\begin{equation}
\delta G^0{}_0=-\delta \rho -\delta \rho _g
\end{equation}
takes the form

\begin{eqnarray}
&&2a^{-2}\left( 3{\cal H}\left( {\cal H}\phi +\psi ^{\prime }\right) -\nabla
^2\psi \right)  \nonumber \\
&&+\left( \frac{3\left( 1-8\gamma \right) }{16\beta \gamma }+\allowbreak
\frac{12\alpha +5\beta -16\gamma \left( 3\alpha +\beta \right) }{8\beta
\gamma }\left( 1+v\right) \rho _b+\allowbreak \frac{\left( 3\alpha +\beta
\right) \left( 4\alpha +\beta \right) }{4\gamma \beta }\left( 1+v\right)
^2\rho _b^2\right) \psi  \nonumber \\
&=&-\frac{1+r+v}{3\gamma }\rho _b\delta -\frac{12\alpha +5\beta -72\alpha
\gamma -20\beta \gamma }{8\beta \gamma }\left( 1+v\right) \rho _b\delta -%
\frac{\left( 3\alpha +\beta \right) \left( 4\alpha +\beta \right) }{2\beta
\gamma }\left( 1+v\right) ^2\rho _b^2\delta ,
\end{eqnarray}
where the growth of the baryonic matter density perturbation $\delta
:=\delta \rho _b/\rho _b$, ${\cal H}:=a^{\prime }/a=aH$, prime denotes
derivative with respect to the conformal time $\eta $.

The equation
\begin{equation}
\delta G^i{}_j=\left( \delta p_r+\delta p_g\right) \delta ^i{}_j,
\end{equation}
reads
\begin{eqnarray}
&&-2a^{-2}\{\left[ \left( 2{\cal H}^{\prime }+{\cal H}^2\right) \phi +{\cal H%
}\phi ^{\prime }+\psi ^{\prime \prime }+2{\cal H}\psi ^{\prime }+\frac 12%
\nabla ^2\left( \phi -\psi \right) \right] \delta ^i{}_j-\frac 12\partial
_i\partial _j\left( \phi -\psi \right) \}+2\left( f^2\psi +f\xi \right)
\delta ^i{}_j  \nonumber \\
&=&\{\frac 13r\rho _b\delta -\frac{\left( 8\gamma +1\right) \left(
1+r+v\right) }{24\gamma }\rho _b\delta +\allowbreak \frac{8\beta \gamma
-3\left( 4\alpha +\beta \right) }{48\beta \gamma }\left( 1+v\right) \rho
_b\delta  \nonumber \\
&&-\frac{\left( 3\alpha +\beta \right) \left( 4\alpha +\beta \right) }{%
12\beta \gamma }\left( 1+v\right) ^2\rho _b^2\delta \}\delta ^i{}_j,
\end{eqnarray}
where $i=1,2,3$.

The equation
\begin{eqnarray}
G^0{}_i &=&R^0{}_i  \nonumber \\
&=&T^0{}_i-4\alpha R{}^0{}_iR-\beta \left( 2R{}^{\rho 0}{}R{}_{\rho
i}+2R{}^{\rho \sigma }{}R{}^0{}_{\rho i\sigma }\right)  \nonumber \\
&&+\gamma \left( 4e_I{}^0\partial _\nu \left( e{}^{I\lambda }T{}_{i\lambda
}{}^\nu \right) -4e{}^K{}_\tau T{}_i{}^{0\nu }\partial _\nu e_K{}^\tau
-4T{}^{\lambda 0}{}_\tau T{}{}_{\lambda i}{}^\tau \right) ,
\end{eqnarray}
yields
\begin{eqnarray}
&&2a^{-2}\left[ {\cal H}\phi +\psi ^{\prime }\right] _{,i}-a^{-1}\left(
8\gamma -1-4\alpha \left( 1+r+v\right) \rho _b\right) \psi _{,i}^{\prime
}-2a^{-1}\left( 1+\frac 43r+v\right) \rho _bV_{,i}  \nonumber \\
&&-4\beta a^{-1}\{\frac{\left( 8\gamma -1\right) \left( 4\gamma +1\right) }{%
16\beta \gamma }-\frac{12\alpha +5\beta -8\gamma \left( 3\alpha +2\beta
\right) }{24\beta \gamma }\left( 1+v\right) \rho _b  \nonumber \\
&&-\allowbreak \frac{\left( 3\alpha +\beta \right) \left( 4\alpha +\beta
\right) }{12\gamma \beta }\left( 1+v\right) ^2\rho _b^2\}{\cal H}\phi
_{,i}{}+8\beta a^{-1}{}{\cal H}f^2\psi _{,i}+8{}a^{-1}{}{\cal H}f\xi _{,i}
\nonumber \\
&=&0.
\end{eqnarray}

In commoving orthogonal coordinates, the three-velocity of baryonic matter
vanishes, $V_b^i=0$ \cite{20}. For the potential $V$ of the three-velocity
field of the dark matter, the perturbed conservation law
\begin{equation}
\delta \left( \nabla _\mu T^\mu {}_\nu +\nabla _\mu T_g{}^\mu {}_\nu \right)
=0,
\end{equation}
leads to the equation
\begin{equation}
\stackrel{\bf \cdot }{V}_{,i}+\left( \frac{\frac 43\stackrel{\cdot }{r}+%
\stackrel{\cdot }{v}}{1+\frac 43r+v}+H\right) V_{,i}=0,
\end{equation}
 when $\nu =i$.

In the case (15) and (16), using (67) and (69) we obtain the perturbation of
the equation (3)
\begin{equation}
2f\xi =\left( \frac{1+r+v}{24\gamma }+\allowbreak \frac{3\left( 4\alpha
+\beta \right) -16\gamma \left( 3\alpha +\beta \right) }{48\beta \gamma }%
\left( 1+v\right) +\allowbreak \frac{\left( 3\alpha +\beta \right) \left(
4\alpha +\beta \right) }{12\gamma \beta }\left( 1+v\right) ^2\rho _b\right)
\rho _b\delta .
\end{equation}

In the Fourier space $k$, from (77) and (79) we obtain the equations of $%
\phi $ and $\psi $,
\begin{eqnarray}
&&2a^{-2}\left( 3{\cal H}\left( {\cal H}\phi +\psi ^{\prime }\right)
+k^2\psi \right)   \nonumber \\
&&+\left( \frac{3\left( 1-8\gamma \right) }{16\beta \gamma }+\allowbreak
\frac{12\alpha +5\beta -16\gamma \left( 3\alpha +\beta \right) }{8\beta
\gamma }\left( 1+v\right) \rho _b+\left( 3\alpha +\beta \right) \allowbreak
\frac{4\alpha +\beta }{4\gamma \beta }\left( 1+v\right) ^2\rho _b^2\right)
\psi   \nonumber \\
&=&-\frac{1+r+v}{3\gamma }\rho _b\delta -\frac{12\alpha +5\beta -72\alpha
\gamma -20\beta \gamma }{8\beta \gamma }\left( 1+v\right) \rho _b\delta -%
\frac{\left( 3\alpha +\beta \right) \left( 4\alpha +\beta \right) }{2\beta
\gamma }\left( 1+v\right) ^2\rho _b^2\delta ,
\end{eqnarray}
and
\begin{eqnarray}
&&-2a^{-2}\{\left[ \left( 2{\cal H}^{\prime }+{\cal H}^2\right) \phi +{\cal H%
}\phi ^{\prime }+\psi ^{\prime \prime }+2{\cal H}\psi ^{\prime }-\frac 12%
k^2\left( \phi -\psi \right) \right] \delta ^i{}_j-\frac 12\partial
_i\partial _j\left( \phi -\psi \right) \}+2\left( f^2\psi +f\xi \right)
\delta ^i{}_j  \nonumber \\
&=&\{\frac 13r\rho _b\delta -\frac{\left( 8\gamma +1\right) \left(
1+r+v\right) }{24\gamma }\rho _b\delta +\allowbreak \frac{8\beta \gamma
-3\left( 4\alpha +\beta \right) }{48\beta \gamma }\left( 1+v\right) \rho
_b\delta -\frac{\left( 3\alpha +\beta \right) \left( 4\alpha +\beta \right)
}{12\beta \gamma }\left( 1+v\right) ^2\rho _b^2\delta \}\delta ^i{}_j.
\end{eqnarray}

When $i\neq j$, (86) leads to
\begin{equation}
\phi =\psi ,
\end{equation}
agreeing with GR but in contrast to $f(R)$ theory \cite{21}. Then we have
the equations of $\psi $:

\begin{eqnarray}
&&2a^{-2}\left( 3{\cal H}\left( {\cal H}\psi +\psi ^{\prime }\right)
+k^2\psi \right)  \nonumber \\
&&+\left( \frac{3\left( 1-8\gamma \right) }{16\beta \gamma }+\allowbreak
\frac{12\alpha +5\beta -16\gamma \left( 3\alpha +\beta \right) }{8\beta
\gamma }\left( 1+v\right) \rho _b+\left( 3\alpha +\beta \right) \allowbreak
\frac{4\alpha +\beta }{4\gamma \beta }\left( 1+v\right) ^2\rho _b^2\right)
\psi  \nonumber \\
&=&-\frac{1+r+v}{3\gamma }\rho _b\delta -\frac{12\alpha +5\beta -72\alpha
\gamma -20\beta \gamma }{8\beta \gamma }\left( 1+v\right) \rho _b\delta -%
\frac{\left( 3\alpha +\beta \right) \left( 4\alpha +\beta \right) }{2\beta
\gamma }\left( 1+v\right) ^2\rho _b^2\delta ,
\end{eqnarray}
\begin{eqnarray}
&&-2a^{-2}\left[ \left( 2{\cal H}^{\prime }+{\cal H}^2\right) \psi +\psi
^{\prime \prime }+3{\cal H}\psi ^{\prime }\right] +2\left( f^2\psi +f\xi
\right)  \nonumber \\
&=&\frac 13r\rho _b\delta -\frac{\left( 8\gamma +1\right) \left(
1+r+v\right) }{24\gamma }\rho _b\delta +\allowbreak \frac{8\beta \gamma
-3\left( 4\alpha +\beta \right) }{48\beta \gamma }\left( 1+v\right) \rho
_b\delta  \nonumber \\
&&-\frac{\left( 3\alpha +\beta \right) \left( 4\alpha +\beta \right) }{%
12\beta \gamma }\left( 1+v\right) ^2\rho _b^2\delta .
\end{eqnarray}

One of the methods to measure the cosmic growth rate is redshift-space
distortion that appears in clustering pattern of galaxies in galaxy redshift
surveys. In order to confront the models with galaxy clustering surveys, we
are interested in the modes deep inside the Hubble radius. In this case we
can employ the quasistatic approximation on sub-horizon scales, under which,
$\partial /\partial \eta \sim {\cal H}\ll k$. Then the perturbation
equations (88), (89) give
\begin{eqnarray}
&&\left( 2a^{-2}k^2-\frac{3\left( 1-8\gamma \right) }{16\beta \gamma }%
-\allowbreak \frac{12\alpha +5\beta -16\gamma \left( 3\alpha +\beta \right)
}{8\beta \gamma }\left( 1+v\right) \rho _b-\left( 3\alpha +\beta \right)
\allowbreak \frac{4\alpha +\beta }{4\gamma \beta }\left( 1+v\right) ^2\rho
_b^2\right) \psi   \nonumber \\
&=&\left( \frac{1+r+v}{3\gamma }\rho _b+\frac{12\alpha +5\beta -72\alpha
\gamma -20\beta \gamma }{8\beta \gamma }\left( 1+v\right) \rho _b+\frac{%
\left( 3\alpha +\beta \right) \left( 4\alpha +\beta \right) }{2\beta \gamma }%
\left( 1+v\right) ^2\rho _b^2\right) \delta ,
\end{eqnarray}
and

\begin{eqnarray}
&&-2a^{-2}\left[ \left( 2{\cal H}^{\prime }+{\cal H}^2\right) \psi +\psi
^{\prime \prime }+3{\cal H}\psi ^{\prime }\right] +2\left( f^2\psi +f\xi
\right)  \nonumber \\
&=&\frac 13r\rho _b\delta -\frac{\left( 8\gamma +1\right) \left(
1+r+v\right) }{24\gamma }\rho _b\delta +\allowbreak \frac{8\beta \gamma
-3\left( 4\alpha +\beta \right) }{48\beta \gamma }\left( 1+v\right) \rho
_b\delta  \nonumber \\
&&-\frac{\left( 3\alpha +\beta \right) \left( 4\alpha +\beta \right) }{%
12\beta \gamma }\left( 1+v\right) ^2\rho _b^2\delta .
\end{eqnarray}

The equation (90) gives the expression of gravitational potential $\psi $.
In the case $\alpha =-\frac \beta 2$, if
\begin{equation}
a^{-2}k^2\gg \rho _b,\left| \alpha \rho _b\right| \gg 1,
\end{equation}
it reduces to the Poisson equation
\begin{equation}
\frac{k^2}{a^2}\psi =-4\pi G_{eff}\rho _b\delta ,
\end{equation}
where
\begin{equation}
G_{eff}=\frac 1{16\pi \gamma }\left( 1+v\right) ^2\alpha \rho _b
\end{equation}
is the effective gravitational coupling constant. In the framework of GR, $%
G_{eff}$ is equivalent to the gravitational constant $G=1$.

\subsection{ Equation of the structure growth and its solution}

Since different theoretical models can achieve the same expansion history of
universe, several methods should be used to discriminate the different
models. The study on the growth of matter density perturbations may become
the useful tool due to that theories with the same expansion history can
have a different cosmic growth history. The perturbation quantities can be
easily related to the cosmic observations \cite{22}.

Using (59), the equations (8) and (9) can be rewritten as
\begin{eqnarray}
H^2 &=&\frac 13\left( \rho _b+\rho _{other}\right) , \\
\stackrel{\cdot }{H} &=&-\frac 12\left( \rho _b+\rho
_{other}+p_{other}\right) ,
\end{eqnarray}
where
\begin{eqnarray}
\rho _{other} &=&\rho _r+\rho _d+\rho _g,  \nonumber \\
p_{other} &=&p_r+p_g.
\end{eqnarray}
We introduce the perturbations of $\rho _b$, $\rho _{other}$, $p_{other}$
and $H$ \cite{23}:
\begin{eqnarray*}
\rho _b &\longrightarrow &\left( 1+\delta \right) \rho _b,\rho
_{other}\longrightarrow \rho _{other}+\delta \rho _{other}, \\
p_{other} &\longrightarrow &p_{other}+\delta p_{other},H\longrightarrow
H+\delta H,
\end{eqnarray*}
with
\begin{equation}
\delta H\equiv \frac 1{3a}\nabla \cdot {\bf u},\;\;{\bf u}=\nabla V.
\end{equation}
Following the approach of \cite{23} and \cite{19}, using (95), (96) and the
perturbed conservation law
\begin{eqnarray}
\delta \left( \nabla _\mu T^\mu {}_\nu +\nabla _\mu T_g{}^\mu {}_\nu \right)
=0,
\end{eqnarray}
we obtain the equation for the growth of the baryonic matter density
perturbation $\delta $:
\begin{eqnarray}
&&\rho _b\stackrel{\cdot \cdot }{\delta }+\frac{\stackrel{\cdot }{\rho }%
_b\rho _{other}-\rho _b\stackrel{\cdot }{\rho }_{other}+\stackrel{\cdot }{%
\rho }_bp_{other}-\rho _b\stackrel{\cdot }{p}_{other}}{\rho _b+\rho
_{other}+p_{other}}\stackrel{\cdot }{\delta }  \nonumber \\
&&-\left( \stackrel{\cdot \cdot }{\rho }_{other}-\frac 32\rho _b\rho
_b-3\left( \rho _{other}+p_{other}\right) \rho _b-\frac 32\left( \rho
_{other}+p_{other}\right) ^2\right) \delta   \nonumber \\
&&+\frac{2\stackrel{\cdot }{\rho }_b+2\stackrel{\cdot }{\rho }_{other}}{\rho
_b+\rho _{other}+p_{other}}\stackrel{\cdot }{p}_{other}\delta +\frac{2%
\stackrel{\cdot }{\rho }_b+2\stackrel{\cdot }{\rho }_{other}}{\rho _b+\rho
_{other}+p_{other}}\stackrel{\cdot }{\rho }_{other}\delta   \nonumber \\
&&-\frac{\left( \stackrel{\cdot }{\rho }_b+\stackrel{\cdot }{\rho }%
_{other}\right) ^2+\left( \stackrel{\cdot }{\rho }_b+\stackrel{\cdot }{\rho }%
_{other}\right) \stackrel{\cdot }{p}_{other}}{\left( \rho _b+\rho
_{other}+p_{other}\right) ^2}\left( \rho _{other}+p_{other}\right) \delta
\nonumber \\
&&+\stackrel{\cdot \cdot }{\delta \rho }_{other}-\frac{2\stackrel{\cdot }{%
\rho }_b+2\stackrel{\cdot }{\rho }_{other}+\stackrel{\cdot }{p}_{other}}{%
\rho _b+\rho _{other}+p_{other}}\stackrel{\cdot }{\delta \rho }_{other}-%
\frac{\stackrel{\cdot }{\rho }_b+\stackrel{\cdot }{\rho }_{other}}{\rho
_b+\rho _{other}+p_{other}}\stackrel{\cdot }{\delta p}_{other}  \nonumber \\
&&-\left( 3\left( \rho _b+\rho _{other}+p_{other}\right) -\frac{\left(
\stackrel{\cdot }{\rho }_b+\stackrel{\cdot }{\rho }_{other}\right) ^2}{%
\left( \rho _b+\rho _{other}+p_{other}\right) ^2}-\frac{\left( \stackrel{%
\cdot }{\rho }_b+\stackrel{\cdot }{\rho }_{other}\right) \stackrel{\cdot }{p}%
_{other}}{\left( \rho _b+\rho _{other}+p_{other}\right) ^2}\right) \left(
\delta \rho _{other}+\delta p_{other}\right)   \nonumber \\
&=&0.
\end{eqnarray}

Up to now, the complete set of equations that describes the general linear
perturbations has been obtained. It provides enough information about the
behaviors of the perturbation  and can be compared with the results of the $%
\Lambda $CDM model.

(71), (74) and (75) yield
\begin{eqnarray}
\stackrel{\cdot }{\rho }_b &=&-3H\rho _b,\stackrel{\cdot }{\rho }_r=-4H\rho
_r,\stackrel{\cdot }{\rho }_d=-nH\rho _d,  \nonumber \\
\stackrel{\cdot }{p}_r &=&-\frac 43H\rho _r,  \nonumber \\
\stackrel{\cdot }{r} &=&-rH,\stackrel{\cdot }{v}=\left( 3-n\right) vH,
\end{eqnarray}
and then (97), (22) and (23) give
\begin{eqnarray}
\rho _{other} &=&\frac{3\left( 8\gamma -1\right) ^2}{32\beta \gamma }+\left(
r+v-\frac{\left( 8\gamma -1\right) \left( 1+r+v\right) }{8\gamma }%
\allowbreak -\frac{3\left( 8\gamma -1\right) \left( 4\alpha +\beta \right) }{%
16\beta \gamma }\left( 1+v\right) \right) \rho _b  \nonumber \\
&&+\allowbreak \frac{\left( 3\alpha +\beta \right) \left( 4\alpha +\beta
\right) }{8\beta \gamma }\left( 1+v\right) ^2\rho _b^2, \\
p_{other} &=&\frac{64\gamma ^2-1}{32\beta \gamma }+\left( \frac 13r-\frac{%
\left( 8\gamma +1\right) \left( 1+r+v\right) }{24\gamma }+\allowbreak \frac{%
8\beta \gamma -3\left( 4\alpha +\beta \right) }{48\beta \gamma }\left(
1+v\right) \right) \rho _b  \nonumber \\
&&-\frac{\left( 3\alpha +\beta \right) \left( 4\alpha +\beta \right) }{%
24\beta \gamma }\left( 1+v\right) ^2\rho _b^2,
\end{eqnarray}
Using (102) and (103), by straightforward and tedious calculations, the equation (100) can be written as
\begin{equation}
\left( 1+r+v+A+\allowbreak D\rho _b\right) \rho _b\stackrel{\cdot \cdot }{%
\delta }+M\left( r,v,\rho _b\right) H\stackrel{\cdot }{\delta }+N\left(
r,v,\rho _b\right) H^2\delta +Q\left( r,v,\rho _b\right) \delta =0,
\end{equation}
where $A$, $D$, $M\left( r,v,\rho _b\right) $, $N\left( r,v,\rho _b\right) $%
, and $Q\left( r,v,\rho _b\right) $ are given in Appendix.

Supposing
\begin{equation}
n=3,
\end{equation}
in the case (31) and (92), i.e. when $\rho _d\propto a^{-3}$, $\beta
=-2\alpha $, and $\left| \alpha \rho _b\right| \gg 1$, the equation (104)
becomes
\begin{equation}
\allowbreak \stackrel{\cdot \cdot }{\delta }-\allowbreak 22H\stackrel{\cdot
}{\delta }+3H^2\delta =0.
\end{equation}
Introduce the logarithmic time variable
\begin{equation}
N=\ln a.
\end{equation}
(106) takes the form
\[
\allowbreak \frac{d^2\delta }{dN^2}-\allowbreak 23\frac{d\delta }{dN}%
+3\delta =0,
\]
and gives the solution
\begin{eqnarray}
\delta  &=&\delta _{0+}a^{\frac 12\left( 23+\sqrt{517}\right) }+\delta
_{0-}a^{\frac 12\left( 23-\sqrt{517}\right) }  \nonumber \\
&\approx &\delta _{0+}a^{22.\,869}+\delta _{0-}a^{0.\,13118},
\end{eqnarray}
which can be compared with the result in GR\cite{24}.

\section{Conclusions}

A cosmology of Poincar$\acute{e}$ gauge theory has been developed. We focus on the case including a pseudoscalar scalar torsion function $f$ as suggested by
Baekler, Hehl and Nester \cite{25}. The gravitational field equation and the two-family cosmological equations has been obtained in Ref. \cite{12}.
In this paper, we focus on studying the second family cosmological equations corresponding to the pseudoscalar torsion function. It is found that although we do not introduce a cosmological constant in the action it automatically emerges in the derivation of the cosmological equations and then is endowed with intrinsic character. It is nothing but the intrinsic torsion and curvature of the vacuum universe. The dark energy is identified with the geometry of the spacetime.
Now we are returning to the original idea of Einstein and Wheeler: gravity is a geometry \cite{26}. The cosmological constant puzzle and the
coincidence and fine tuning problem are solved naturally. The point is that the dark energy is the functions of the density and pressure of the cosmic fluid and includes the cosmological constant but can not be identified with it. The analytic expressions of the state equation and the density parameters of the matter
and the geometric dark energy are derived and used to determine the values of $\alpha $ , $\beta $ and $\gamma $. Then a theoretical value of the
cosmological constant is computed and compared with the observed datum. An analytic integral of the cosmological equation is obtained and used to
evaluate the age of the universe which can be compared with observed data. The full equations of linear cosmological perturbations and the solutions
are obtained. In addition, the behavior of perturbations for the sub-horizon modes relevant to large-scale structures is discussed. This model can be
distinguished from others by considering the evolution of matter perturbations and gravitational potentials.

 \textbf{\ Acknowledgments }
We thank the anonymous referee for his/her very instructive comments, which improve our paper greatly. The research work is supported by   the National Natural Science Foundation of China (11205078).

\appendix

\section{$\text{The growth of structures in linear perturbation theory}$}

In the following, we give the derivation of the equation (104):

Letting
\begin{eqnarray}
A &=&-\frac{\left( 8\gamma -1\right) \left( 1+r+v\right) }{8\gamma }%
\allowbreak -\frac{3\left( 8\gamma -1\right) \left( 4\alpha +\beta \right) }{%
16\beta \gamma }\left( 1+v\right) ,  \nonumber \\
B &=&-\frac{\left( 8\gamma +1\right) \left( 1+r+v\right) }{24\gamma }%
+\allowbreak \frac{8\beta \gamma -3\left( 4\alpha +\beta \right) }{48\beta
\gamma }\left( 1+v\right) ,  \nonumber \\
D &=&\allowbreak \frac{\left( 3\alpha +\beta \right) \left( 4\alpha +\beta
\right) }{4\beta \gamma }\left( 1+v\right) ^2.
\end{eqnarray}
\begin{eqnarray}
E &=&-\frac{\left( 8\gamma -1\right) \left( \left( 3-n\right) v-r\right) }{%
8\gamma }\allowbreak -\frac{3\left( 8\gamma -1\right) \left( 4\alpha +\beta
\right) }{16\beta \gamma }\left( 3-n\right) v,  \nonumber \\
F &=&-\frac{\left( 8\gamma +1\right) \left( \left( 3-n\right) v-r\right) }{%
24\gamma }+\allowbreak \frac{8\beta \gamma -3\left( 4\alpha +\beta \right) }{%
48\beta \gamma }\left( 3-n\right) v,  \nonumber \\
K &=&\allowbreak \frac{\left( 3\alpha +\beta \right) \left( 4\alpha +\beta
\right) }{2\beta \gamma }\left( 1+v\right) \left( 3-n\right) v,  \nonumber \\
L &=&\frac{\left( 8\gamma -1\right) \left( \left( 3-n\right) ^2v+r\right) }{%
8\gamma }\allowbreak +\frac{3\left( 8\gamma -1\right) \left( 4\alpha +\beta
\right) }{16\beta \gamma }\left( 3-n\right) ^2v,
\end{eqnarray}
(89), (90) and (91) give
\begin{eqnarray}
\rho _{other} &=&\frac{3\left( 1-8\gamma \right) ^2}{32\beta \gamma }+\left(
r+v+A\right) \rho _b+\frac D2\rho _b^2,  \nonumber \\
p_{other} &=&\frac{64\gamma ^2-1}{32\beta \gamma }+\left( \frac 13r+B\right)
\rho _b\allowbreak -\frac D6\rho _b^2.
\end{eqnarray}
Then we compute
\begin{eqnarray}
\stackrel{\cdot }{\rho }_{other} &=&-\left( vn+4r\right) H\rho _b+EH\rho _b+%
\frac K2H\rho _b^2,  \nonumber \\
\stackrel{\cdot }{p}_{other} &=&-\frac 43rH\rho _b+FH\rho _b\allowbreak -%
\frac K6H\rho _b^2,
\end{eqnarray}
\begin{eqnarray}
\stackrel{\cdot \cdot }{\rho }_{other} &=&\left( n^2v+16r-L-3E\right)
H^2\rho _b  \nonumber \\
&&-\frac 12\left( n+3\right) KH^2\rho _b^2+\frac K2\frac{\left( 3-n\right) v%
}{\left( 1+v\right) }H^2\rho _b^2  \nonumber \\
&&+\left( E-\left( vn+4r\right) \right) \stackrel{\cdot }{H}\rho _b+\frac K2%
\stackrel{\cdot }{H}\rho _b^2,
\end{eqnarray}
\begin{eqnarray}
\delta \rho _{other} &=&\left( r+v+A\right) \rho _b\delta +\allowbreak D\rho
_b^2\delta ,  \nonumber \\
\delta p_{other} &=&\left( \frac 13r+B\right) \rho _b\delta -\frac D3\rho
_b^2\delta ,
\end{eqnarray}
\begin{eqnarray}
\stackrel{\cdot }{\delta \rho }_{other} &=&-\left( vn+4r\right) H\rho
_b\delta +\left( r+v\right) \rho _b\stackrel{\cdot }{\delta }+A\rho _b%
\stackrel{\cdot }{\delta }+\allowbreak D\rho _b^2\stackrel{\cdot }{\delta }%
-\left( 3A-E\right) H\rho _b\delta -\left( \allowbreak 2D-\allowbreak
K\right) H\rho _b^2\delta ,  \nonumber \\
\stackrel{\cdot }{\delta p}_{other} &=&-\frac 43rH\rho _b\delta +\frac 13%
r\rho _b\stackrel{\cdot }{\delta }+B\rho _b\stackrel{\cdot }{\delta }-\frac D%
3\rho _b^2\stackrel{\cdot }{\delta }-\left( 3B-F\right) H\rho _b\delta
+\left( 2D-\frac K3\right) H\rho _b^2\delta ,
\end{eqnarray}
and
\begin{eqnarray}
\stackrel{\cdot \cdot }{\delta \rho }_{other} &=&\left( r+v+\left(
A+\allowbreak D\rho _b\right) \right) \rho _b\stackrel{\cdot \cdot }{\delta }
\nonumber \\
&&-2\left[ \left( nv+4r\right) -2\left( E-3A+\allowbreak \left(
K-6\allowbreak D\right) \rho _b\right) \right] H\rho _b\stackrel{\cdot }{%
\delta }  \nonumber \\
&&+\left[ n^2v+16r+9A-6E-L+\left( 36\allowbreak D-12\allowbreak
K+\allowbreak \frac{1+2v}{1+v}K\left( 3-n\right) \right) \rho _b\right]
H^2\rho _b\delta   \nonumber \\
&&+\left( E-3A-\left( vn+4r\right) +\left( \allowbreak K-6\allowbreak
D\right) \rho _b\right) \stackrel{\cdot }{H}\rho _b\delta .
\end{eqnarray}
Substituting these into (100) yields (104):
\begin{equation}
\left( 1+r+v+A+\allowbreak D\rho _b\right) \rho _b\stackrel{\cdot \cdot }{%
\delta }+M\left( r,v,\rho _b\right) H\stackrel{\cdot }{\delta }+N\left(
r,v,\rho _b\right) H^2\delta +Q\left( r,v,\rho _b\right) \delta =0,
\end{equation}
where
\begin{eqnarray}
M\left( r,v,\rho _b\right)  &=&-\left( 3+4r+2vn-3v+9A-3B-4E\right) \rho
_b+\left( 4K-22\allowbreak D\right) \rho _b^2  \nonumber \\
&&+\frac{\left( -\left( 1+r+v+A\right) \rho _b+\allowbreak D\rho _b^2\right)
\left( -vn-\allowbreak \frac{16}3r+E+F\allowbreak -3\right) }{\frac{\left(
8\gamma -1\right) \left( 16\gamma -1\right) }{16\beta \gamma }+\left( 1+%
\frac 43r+v+A+B\right) \rho _b\allowbreak +\frac 13D\rho _b^2}\rho _b
\nonumber \\
&&+\frac{-\frac 13\left( 1+r+v+A\right) \rho _b+\frac 13\allowbreak D\rho
_b^2}{\frac{\left( 8\gamma -1\right) \left( 16\gamma -1\right) }{16\beta
\gamma }+\left( 1+\frac 43r+v+A+B\right) \rho _b\allowbreak +\frac 13D\rho
_b^2}K\rho _b^2,
\end{eqnarray}
\begin{eqnarray}
N\left( r,v,\rho _b\right)  &=&-\frac{\stackrel{\cdot }{\frac{3\left(
8\gamma -1\right) \left( 16\gamma -1\right) }{16\beta \gamma }}+3A\rho
_b+\left( \allowbreak D-\frac{\allowbreak K}2\right) \rho _b^2}{\frac{\left(
8\gamma -1\right) \left( 16\gamma -1\right) }{16\beta \gamma }+\left( 1+%
\frac 43r+v+A+B\right) \rho _b\allowbreak +\frac 13D\rho _b^2}\left(
3+\allowbreak \frac{16}3r+vn-E-F\allowbreak \right) \rho _b  \nonumber \\
&&+\left( \allowbreak -9B+36D\rho _b-3E+\frac 12\left( n-19\right) K\right)
\rho _b+\allowbreak \frac{\allowbreak 2+3v}{2\left( 1+v\right) }\left(
3-n\right) K\rho _b^2  \nonumber \\
&&+\frac{\stackrel{\cdot }{\frac{\left( 8\gamma -1\right) \left( 16\gamma
-1\right) }{16\beta \gamma }}+A\rho _b+\frac 13\left( \allowbreak D-\frac{%
\allowbreak K}2\right) \rho _b^2}{\frac{\left( 8\gamma -1\right) \left(
16\gamma -1\right) }{16\beta \gamma }+\left( 1+\frac 43r+v+A+B\right) \rho
_b\allowbreak +\frac 13D\rho _b^2}K\rho _b^2,
\end{eqnarray}
and
\begin{eqnarray}
Q\left( r,v,\rho _b\right)  &=&+\frac 32\left( \frac{\left( 8\gamma
-1\right) \left( 16\gamma -1\right) }{16\beta \gamma }\right) ^2-\frac 32%
\frac{\left( 8\gamma -1\right) \left( 16\gamma -1\right) }{16\beta \gamma }%
\rho _b  \nonumber \\
&&-\left( \frac{\left( 8\gamma -1\right) \left( 16\gamma -1\right) }{16\beta
\gamma }D+\frac 32\left( 1+\frac 43r+v+A+B\right) ^2+\frac 32\left( 1+\frac 4%
3r+v+A+B\right) \right) \rho _b^2  \nonumber \\
&&-\left( 2\left( 1+\frac 43r+v+A+B\right) +\allowbreak \frac 12\right)
D\rho _b^3-\frac 12D^2\rho _b^4  \nonumber \\
&&-3A\stackrel{\cdot }{H}\rho _b\delta +\left( \frac K2-6\allowbreak
D\right) \stackrel{\cdot }{H}\rho _b^2.
\end{eqnarray}


\begin{thebibliography}{*}



\bibitem{1}
Amendola L and Tsujikawa S 2010 {\it Dark Energy} (Cambridge University press);
 Ade P A R, et al (Planck Collaboration) [arXiv:1502.01589];
 Mortonson M J, Weinberg D H, White M [arXiv:1401.0046];
Cai R G, Guo Z K, and Yang T  2016 Phys. Rev. D 93, 043517 [arXiv:1509.06283];
Cai R G  and Wang S J 2016 Phys. Rev. D 93, 023515 [arXiv:1511.00627];
Battye R A, Moss A 2014 Phys. Rev. Lett. 112, 051303 [arXiv:1308.5870];
Zhang J F, Zhao M M, Li Y H, Zhang X 2015 JCAP 04 038 [arXiv:1502.04028];
 Lu J B, Xu Y F, and Wu Y B 2015  Eur. Phys. J. C 75, 473 [arXiv:1503.02439];
 Lu J B, Geng D H, Xu L X, Wu Y B, and Liu M L 2015 JHEP 02, 071 [arXiv:1312.0779];
 Davis T M 2014 General Relativity and Gravitation 46,1731 [1404.7266].


\bibitem{2}
Capozziello S, Harko T, Koivisto T S, Lobo F S N, and Olmo G J 2015 Universe 1(2), 199-238 [arXiv:1508.04641];
Takahashi K, and Yokoyama J 2015 Phys. Rev. D 91, 084060 [arXiv:1503.07412];
Preston A W H, Morris T R 2014 JCAP 09, 017 [1406.5398];
Clifton T, Ferreira P G, Padilla A and Skordis C 2012 Phys. Rept. 513, 1 [arXiv:1106.2476];
Capozziello S and De Laurentis M 2011 Phys. Rept. 509, 167 [arXiv:1108.6266 [gr-qc]]; Bamba K,
Capozziello S, Nojiri S and Odintsov S D [arXiv:1205.3421 [gr-qc]];
 Jaime L G, Patino L and Salgado M arXiv:1206.1642 [gr-qc];
 Nojiri S and Odintsov S D 2011 Phys. Rept. 505, 59
[arXiv:1011.0544 [gr-qc]]; Capozziello S and Faraoni V 2010 Beyond Einstein Gravity (Springer);
Felice A D, Tsujikawa 2010 Living Rev. Rel. 13,3  [arXiv:1002.4928]

\bibitem{3}  Barboza Jr E M, Nunes R C, Abreu E M C, and Neto J A
arXiv:1501.03491v1 [gr-qc].

\bibitem{4}  Multamaki T and Vilja I 2006 Phys. Rev. D 74, 064022; Faraoni V
2004 Phys. Rev. D 70, 081501; Flanagan E 2004 Class. Quant. Grav. 21, 3817;
Fujii Y and Maeda K 2003 {\it The Scalar-Tensor Theory of Gravitation}
(Cambridge Univ. Press, Cambridge); Magnano G and Sokolowski L M 1994 Phys.
Rev. D 50 5039

\bibitem{5}  Capozziello S, De Laurentis M and Luongo O arXiv:1406.6996
[gr-qc]; Jaime L G, Patino L, Salgado M arXiv:1312.5428 [gr-qc]; Jaime L G,
Patino L and Salgado M arXiv:1206.1642 [gr-qc].

\bibitem{6}  Hehl F W, McCrea J D, Mielke E W and Ne'eman Y 1995 Phys. Rep.
258 1; Hehl F W arXiv:1204.3672 [gr-qc]; Puetzfeld D and Obukhov Y N 2014
Phys. Rev. D 90, 085034 [arXiv:1408.5669 [gr-qc]]; Obukhov Y N and Puetzfeld
D 2014 Phys. Rev. D 90, 024004 [arXiv:1405.4003 [gr-qc]]; Ali S A, Cafaro C,
Capozziello S, Corda C 2009 Int. J. Theor. Phys.48, 3426 [arXiv:0907.0934
[gr-qc]]; Wang C-H,  Wu Y-H 2009 Class Quantum Grav. 26, 045016 [arXiv:0807.0069];
 Puetzfeld D 2005 New Astron. Rev.  49, 59 [arXiv:gr-qc/0404119];
  Cai Y F,   Capozziello S,  Laurentis M D, E  Saridakis E N [arXiv:1511.07586].

\bibitem{7}  Capozziello S, Carloni S, Lambiase G, Stornaiolo C and Troisi A
arXiv:gr-qc/0111106; Capozziello S, Cardone V F, Piedipalumbo E, Sereno M,
and Troisi A 2003 Int. J. Mod. Phys. D12, 381 [arXiv:astro-ph/0209610];
Capozziello S, Cianci R, Stornaiolo C, and Vignolo S 2007 Class.Quant.Grav.
24, 6417 [arXiv:0708.3038 [gr-qc]]; Poplawski N J 2010
Phys.Lett.B694:181-185 [arXiv:1007.0587 [astro-ph.CO]], 2012 Phys. Rev. D
85, 107502 [arXiv:1111.4595 [gr-qc]], 2013 Astron. Rev. 8, 108
[arXiv:1106.4859 [gr-qc]]

\bibitem{8}  Yo H-J and Nester J M 1999 Int. J. Mod. Phys. D8, 459
arXiv:gr-qc/9902032; Shie K-F, Nester J M and Yo H-J 2008 Phys. Rev. D 78
023522 [arXiv:0805.3834 [gr-qc]]; Chen H, Ho F.-H, Nester J M, Wang C-H and
Yo H-J 2009 JCAP 0910, 027 arXiv:0908.3323 [gr-qc];  Ho F-H,
Nester J M arXiv:1106.0711 [gr-qc]; Ho F-H, Chen H, Nester J M, and Yo H-J, Chinese J. Phys. 53 (2015) 110109 [arXiv:1512.01202].

\bibitem{9}  Garkun A S, Kudin V I, Minkevich A V, arXiv:1410.0460
[gr-qc];Minkevich A V arXiv:1309.6075 [gr-qc]; Minkevich A V, Garkun A S,
Kudin V I, 2013 JCAP03, 040 (arXiv:1302.2578 [gr-qc]); Garkun A S, Kudin V
I, Minkevich A V and Vasilevsky Y G arXiv:1107.1566 [gr-qc]; Minkevich A V,
arXiv:1102.0620 [gr-qc]; Minkevich A V 2011 Mod. Phys. Lett. A 26, 259
(arXiv:1002.0538 [gr-qc]); Minkevich A V 2009 Phys. Lett. B 678, 423
(arXiv:0902.2860 [gr-qc]); Minkevich A V, Garkun A S and Kudin V I 2007
Class. Quantum Grav. 24, 5835 (arXiv:0706.1157 [gr-qc])

\bibitem{10}  Lu H, Perkins A, Pope C N and Stelle K S 2015 Phys. Rev. Lett.
114, 171601 (arXiv:1502.01028 [hep-th]); Lu H, Perkins A, Pope C N and
Stelle K S, arXiv:1508.00010 [hep-th]; Stelle K S 1977 Phys. Rev. D 16, 953;
Hindawi A, Ovrut B. A, and Waldram D 1996 Phys. Rev. D 53, 5583; Deser S and
Tekin B 2002 Phys. Rev. Lett. 89, 101101 [arXiv: hep-th/0205318]; Barrow J D
and Hervik S 2006 Phys. Rev. D74, 124007 [arXiv:gr-qc/0610013]; Barrow J D
and Middleton J 2007 Phys. Rev. D75, 123515 [arXiv:gr-qc/0702098]

\bibitem{11}  Polchinski J 1998 {\it String Theory }(Cambridge Univ. Press,
Cambridge).

\bibitem{12}  Chee G and Guo Y 2012 Class. Quantum Grav. 29, 235022
[arXiv;1205.5419[gr-qc]]

\bibitem{13}  Starobinsky A A 1980 Phys. Lett. B 91, 99

\bibitem{14}  Tsamparlis M 1979 Phys. Lett. A 75, 27 ; Goenner H F M and
Muller-Hoissen F 1984 Class. Quant. Grav. 1, 651 ; Baekler P Exact solutions
in the Poincar$\acute{e}$ gauge field theory of gravitation, Ph.D. thesis (University
of Cologne, Germany, 1986).

\bibitem{15}  Mortonson M J, D H Weinberg D H and White M arXiv:1401.0046
[astro-ph.CO]

\bibitem{16}  Davis T M arXiv:1404.7266 [astro-ph.CO]

\bibitem{17}  Frieman J A, Turner M S and Huterer D 2008 arXiv:0803.0982
[astro-ph]

\bibitem{18}  Comelli D, Crisostomi M, and Pilo L 2014 Phys. Rev. D 90,
084003 (arXiv:1403.5679 [hep-th]); Lima N A and Liddle A R 2013 Phys. Rev. D
88, 043521 (arXiv:1307.1613 [astro-ph.CO]); Alvarenga F G, de la
Cruz-Dombriz A, Houndjo M J S, Rodrigues M E, and Saez-Gomez D 2013 Phys.
Rev. D 87, 103526 (arXiv:1302.1866 [gr-qc]); Matsumoto J 2013 Phys. Rev.
D87, 104002 (arXiv:1303.6828 [hep-th])

\bibitem{19}  Peter P arXiv:1303.2509 [astro-ph.CO]; Lesgourgues J
arXiv:1302.4640 [astro-ph.CO]; Bernardeau F, Colombi S, Gaztanaga E, and
Scoccimarro R 2002 Phys. Rept. 367, 1 (arXiv:astro-ph/0112551); Ma C-P and
Bertschinger E 1995 Astrophys. J. 455, 7 (arXiv:astro-ph/9506072); Mukhanov
V F, Feldman H A and Brandenberger R H 1992 Phys. Rept. 215, 203; Bardeen J
M 1980 Phys. Rev. D22 1882

\bibitem{20}  Malik K A and Wands D 2009 Phys. Rept. 475, 1 (arXiv:0809.4944
[astro-ph])

\bibitem{21}  Weinberg D. H., Mortonson M. J., Eisenstein D. J., Hirata C.,
Riess A. G., and Rozo E. arXiv:1201.2434 [astro-ph.CO]

\bibitem{22}  Zhao G B, Li H, Linder E V, Koyama K, Bacon D J and Zhang X
2012 Phys. Rev. D 85, 123546 (arXiv:1109.1846 [astro-ph]); Bean R, Bernat D,
Pogosian L, Silvestri A and Trodden M 2007 Phys. Rev. D75, 064020
(astro-ph/0611321); Koivisto T and Kurki-Suonio H 2006 Class. Quant. Grav.
23, 2355 (astro-ph/0509422); Hwang J and Noh H 2002 Phys. Rev. D65, 023512
[astro-ph/0102005]

\bibitem{23}  Lifshitz E M 1946 J. Phys. (USSR) 10, 116; Bonnor W B 1957
Mon. Not. R. Astron. Soc. 117, 104; Lifshitz E M and Khalatnikov I M 1963
Adv. Phys. 12, 185; Nariai H 1969 Prog. Theor. Phys. 41, 686; Noh H and
Hwang J 2004 Phys. Rev. D 69, 104011; Hwang J and Noh H 2005 Phys.Rev. D71,
063536 (arXiv:gr-qc/0412126); Hwang J and Noh H 2005 Phys.Rev. D72, 044011
(arXiv:gr-qc/0412128); Hwang J and Noh H 2005 Phys.Rev. D72, 044012
(arXiv:gr-qc/0412129)

\bibitem{24}  Copeland E J, Sami M and Tsujikawa S 2006 Int. J. Mod. Phys. D
15, 1753 (arXiv:hep-th/0603057)

\bibitem{25}  Baekler P, Hehl F W and Nester J M 2011 Phys. Rev. D 83,
024001 [arXiv:1009.5112 [gr-qc]]

\bibitem{26}  Chiu H-Y and Hoffmann W F 1964 {\it Gravitation and Relativity
}(W A Benjamin Inc New York) Chapter 3, 4; Misner C W, Thorne K S and
Wheeler J A 1973 {\it Gravitation }(W H Freeman and Company San Francisco)



\end{thebibliography}
\end{document}